\documentclass[twocolumn,showpacs,preprintnumbers,prb,aps]{revtex4-2}
\usepackage{amsmath}
\usepackage{amssymb}
\usepackage{graphicx}
\usepackage{xcolor}
\usepackage[caption=false]{subfig}
\usepackage{placeins}
\usepackage[colorlinks=true, linkcolor=blue, citecolor=blue, filecolor=blue, urlcolor=blue,pdfproducer={.},pdfcreator={.}]{hyperref}
\usepackage{cleveref}
\usepackage{blindtext}
\usepackage{comment}
\usepackage{tikz}
\usepackage{verbatim}
\usepackage[squaren, textstyle]{SIunits}
\usepackage[normalem]{ulem}

\definecolor{fb}{RGB}{0,44,83}
\definecolor{unired}{RGB}{205,9,32}

\newcommand{\zit}[1]
       {\cite{#1}}  % choose one of both!
\begin{document}

%Title of paper
\title{Intertwined relaxation processes maintain athermal electron distribution\\ in laser-excited dielectrics}

\author{Nils Brouwer}
%\email{nils.brouwer@xfel.eu}
\affiliation{European XFEL GmbH, Holzkoppel 4, 22869 Schenefeld, Germany}

\author{Steffen Hirtle}
\author{Baerbel Rethfeld}
\email{rethfeld@rptu.de}
\affiliation{Department of Physics and Research Center OPTIMAS, RPTU
Kaiserslautern-Landau, 
67663 Kaiserslautern, Germany}

\date{\today}

\begin{abstract}

We study the relaxation dynamics of laser-excited non-equilibrium electron distributions in the valence- and conduction band of a dielectric. 
We apply Boltzmann collision integrals to trace the influence of different scattering mechanisms on the energy- and particle density of electrons and holes. 
Our results show a two-timescale behavior of the equilibration process: 
Thermalization within each band towards a respective Fermi distribution as well as equilibration of the band-resolved temperatures occur within a few femtoseconds. 
In contrast, the equilibration of the respective chemical potentials, driven by scattering processes involving particle exchange like impact ionization and Auger recombination, requires timescales in the range of hundreds of femtoseconds. 
We evaluate the effect of our model assumptions for distinct material parameters on the extracted specific relaxation times. 
All these timescales, however, strongly increase, when the additional scattering channel with the cold phonon system is considered: Our simulations demonstrate that an athermal non-Fermi electron distribution can be maintained well up to the picosecond range. 

\end{abstract}


\maketitle

\section{Introduction}
\label{sec:intro}

Ultrashort laser pulses are relevant in numerous applications due to their ability to deposit energy in strongly confined spatial and temporal scales, {\em e.g.}, in laser--processing~\cite{BaeuerleBuch11}, nanosurgery~\cite{Vogel2005}, as well as magnetic switching \cite{Mangin2014NatureMaterials}. 
The very high intensities achieved with ultrashort laser pulses also enable energy deposition in  
initially transparent materials such as semiconductors and dielectrics by multi--photon absorption across the band gap~\cite{Keldysh1965b,Balling2013}.
The interaction of ultrashort laser pulses with solids is also interesting from a more fundamental point  
of view, since the pulse durations are comparable to the intrinsic timescales of fundamental scattering processes within the electronic 
degrees of freedom
and the lattice \zit{Koopmans2010,Bauer2015,Winkler2017}.

In the search for a theoretical description of laser-induced processes on multiple timescales, a short pulse duration allows the laser excitation to be separated from the following relaxation processes~\cite{shugaev2016fundamentals,Rethfeld2017}: At first the laser pulse excites the electrons, which may lead to a state of non-equilibrium inside 
as well as in between  
electronic bands, in which a common electron temperature is not defined. 
In addition, there is a non-equilibrium between the electronic and phononic systems, since only the electrons can be directly excited by optical lasers while the lattice remains cold. 
Several theoretical methods have been developed to describe laser excitation and the following relaxation processes. One very successful method is the two-temperature model (TTM)~\cite{Anisimov1974}, in which the electronic and phononic systems are each assumed to be described by a specific temperature.
Several extensions have been proposed to account, {\em e.g.}, for multiple electron bands in semiconductors \cite{vanDriel1987,Raemer2014} or in noble metals \cite{Ndione2022,Ndione2022b}, spin-resolution in 
itinerant ferromagnets \cite{Beaurepaire1996,Koopmans2000,Mueller2014PRB}, as well as several phonon modes \cite{Waldecker2016}.
However, these temperature-based models inherently assume equilibrium distribution functions (Fermi distributions for electrons, Bose-distributions for phonons) and are therefore unable to capture athermal electron distributions and to describe their thermalization.
An approximate consideration of athermal electrons has been proposed with a separation of the electronic system, where the laser-excited electrons are treated separately from the unaffected electrons \cite{Carpene2006, Tsibidis2018, Uehlein2022}.
Another approach to describe the excitation of solids with ultrashort laser pulses 
is time-dependent density functional theory. It has been applied to the laser excitation of metals in combination with molecular dynamics~\cite{Miyamoto2021} and was used to study the light-matter interaction in dielectrics~\cite{Yamada2019}.
However, these models lack the capability to track electron-electron relaxation.

In this work we apply Boltzmann collision integrals to track the energy-resolved electron and phonon distribution functions in a laser-excited dielectric during and after irradiation with an ultrashort laser pulse of a photon energy much smaller than the band gap~\cite{Kaiser2000,Brouwer2014,brouwer2017}. 
Here, excitation of valence-band electrons is possible by multiphoton-ionization as well as tunneling ionization. 
Further single-photon absorption of electrons may occur in the conduction as well as in the valence band 
through inverse Bremsstrahlung (intraband absorption).  
Electron-electron collisions can redistribute energy within one band (intraband), or may exchange energy as well as particles between both bands (interband). 
The electron systems loose energy due to interaction with the cold lattice, which is described through the interaction of the band-resolved electrons with several phonon modes. 
This approach allows us to study the equilibration of energy inside each considered electron band as well as the interplay of equilibration between conduction band, valence band and phonon modes without assuming an equilibrium distribution within any subsystem.
The model follows our previous works in Refs.~\cite{Brouwer2014,brouwer2017,phd_Brouwer}.

We will briefly recall the applied model in the next section~\ref{sec:theory}, 
and explain the most relevant theoretical aspects for this work. 
In section~\ref{sec:results} we show our results for different assumptions and starting conditions. 
Specifically, section~\ref{sec:electronrelax-wlaser-noeph} deals with the time evolution of the laser-excited electronic energy distributions,  
when the influence of the phonon systems is neglected. 
Distinct features of the transient non-equilibrium electron distribution function, particularly its difference to a thermal 
distribution, are identified and attributed to the considered scattering processes. 
The observed different timescales of characteristic equilibration processes are studied in more detail in section~\ref{sec:nolaser}, where we initialize
excited Fermi distributions for conduction and valence band, respectively, 
and analyze the influence of 
varying excitation strengths and material parameters on the resulting 
thermalization times. 
In section~\ref{sec:electronrelax-wlaser-weph} we trace the temporal evolution of laser-excited electron distribution 
including additionally the interaction with the phonons. 
This leads to cooling of the excited electronic systems, which 
continuously feeds a certain nonequilibrium in the electron distribution lasting 
up to picosecond timescales.
Our results therefore show that the electron-phonon coupling causes the electrons' energy distribution to remain athermal on timescales well beyond their intrinsic thermalization times. 

%--------------------------------------------------------------------
\section{Theory}
\label{sec:theory}
In this work, we apply Boltzmann collision integrals 
to trace the transient distribution functions of electrons and phonons in a dielectric. The considered systems are an electronic valence and conduction band as well as one longitudinal acoustic phonon mode and two longitudinal optical phonon modes. 
Each electronic band is described by an isotropic and parabolic dispersion relation,
with differing effective masses with opposite signs.
The band edges are separated by a band gap.

The change of the electronic energy distribution is given by the Boltzmann equation. 
Since we assume a homogeneous, isotropic material, we neglect spatial and angular 
dependencies. 
The energy distribution $f$ then depends on energy $E$ and time $t$ only, and its changes are given by 
a sum of all considered collision integrals
\begin{equation}
\frac{{\partial}f_\alpha(E,t)}{{\partial}t}
=\sum\limits_{\text{coll}}\left.\frac{\partial f_\alpha(E,t)}{\partial t}\right\vert_{\text{coll}} \, ,
\label{eqn:isobm}
\end{equation}
where the subscript $\alpha$ indicates conduction or valence band, respectively.

Several collision processes, within each band as well as interactions between the bands, 
are considered in our model: 
Electron--electron intraband collisions allow for thermalization within one band. 
Energy is exchanged between the two bands by collisions between electrons of different bands, also termed as electron--hole collisions. 
In addition, Auger recombination and impact ionization allow for particle exchange between both bands.  
During impact ionization, a high-energy electron of the conduction band transfers part of its energy to an electron in the valence band, enabling the latter to overcome the band gap. 
This process is also known as secondary electron ionization and may induce an avalanche process~\zit{Balling2013,Winkler2017,Rethfeld2004a}.
It results in two electrons of low kinetic energy, 
which can, in turn, also
be the initial situation of the opposite process, namely Auger recombination.
The energy absorption from the laser irradiation is described by a photoionization term (including both, multiphoton and tunneling ionization) and by inverse Bremsstrahlung (intraband absorption) described by an electron--phonon--photon collision integral. 

The Boltzmann collision equation for the phonon distribution $s$ of mode $\beta$ reads completely analogous to the one for the electrons~\eqref{eqn:isobm}, namely 
\begin{align}
\label{eqn:phbm}
\frac{{\partial}s_{\beta}(E,t)}{{\partial}t}=\sum\limits_{\text{coll}}\left.\frac{\partial s_{\beta}(E,t)}{\partial t}\right\vert_{\text{coll}} \, .
\end{align}
The phonon modes exchange energy with 
both electronic bands by electron--phonon (e-ph) collisions. 
They also serve as a momentum-reservoir for photon absorption during electron--phonon--photon (e-ph-pt) collisions.
Phonon--phonon collisions are neglected in our approach as they are relevant on longer timescales than considered here \zit{Klett2018,Caruso2022}.

The distinct collision integrals on the right-hand side of Eqs. \eqref{eqn:isobm} and \eqref{eqn:phbm} are derived using first-order perturbation theory and Fermi's golden rule. Our approach is explained in detail in Refs.~\cite{Brouwer2014,brouwer2017}. 
Here, we repeat only a few expressions that will be most relevant for the results shown in this work.

For electron--electron collisions such as Auger recombination and impact ionization, we start with two electrons with initial wave vectors $\mathbf{k}$ and $\mathbf{k_2}$ which transition into states with wave vectors $\mathbf{k_1}$ and $\mathbf{k_3}$. Since both Auger recombination and impact ionization involve 
a transition of an electron from one band to another, the overlap integral of both bands, which accounts for the probability of band transition, must be included.
The overlap integral 
between the Bloch factor $u_{c,\mathbf{k}}$ of the conduction band state with momentum $\mathbf{k}$ and $u_{v,\mathbf{k_1}}$ of the valence band state with momentum $\mathbf{k_1}$ is given by
\begin{equation}
I_{cv}(\mathbf{k},\mathbf{k_1}) = \int_{\text{unit cell}} d\mathbf{r} \,
u_{v,\mathbf{k_1}}^*(\mathbf{r}) u_{c,\mathbf{k}}(\mathbf{r})\,.
\end{equation}
Using $\mathbf{k} \cdot \mathbf{p}$-theory and f-sum rule, it can be approximated by \cite{Ridley1999}
\begin{equation}
\left| I_{cv}(\mathbf{k},\mathbf{k_1}) \right|^2 \approx \frac{\hbar^2}{2 \Delta} \left(\frac{1}{m_c^*} + \frac{1}{m_v^*} \right) \kappa^2 \, ,
\end{equation}
where $\kappa=|\mathbf{k}-\mathbf{k_1}|=|\mathbf{k_2}-\mathbf{k_3}|$ is the exchanged momentum, $\Delta$ is the band gap energy and $m_c^*$ and $m_v^*$ are the effective conduction and valence band masses, respectively. 
The overlap integral introduces a probability for band transition to the electron--electron interaction ($e-e$) and thus directly enters the matrix element for both Auger recombination (Aug) and impact ionization (imp)~\cite{Ridley1999},
\begin{eqnarray}
|M_\text{Aug}|^2 = |M_\text{imp}|^2 &=& |M_{e-e}|^2 \cdot \left| I_{cv}(\kappa) \right|^2 \nonumber\\
&=& \frac{e^4}{\epsilon_0^2 V^2} \frac{\left| I_{cv}(\kappa) \right|^2}{q_0^2 + \kappa^2} \delta_{\mathbf{k}+\mathbf{k}_{2},\mathbf{k}_{1}+\mathbf{k}_{3}}\,,
\end{eqnarray}
with the electron charge $e$, vacuum permittivity $\epsilon_0$, volume of the unit cell $V$ and the inverse screening length $q_0$.

For electron-phonon collisions, we apply two different matrix elements, depending on the considered phonon mode. 
Longitudinal acoustic (LA) phonons are described in the framework of the 
Debye model with a linear dispersion relation $\omega_\text{LA}(q) = c_s \, q$, where $c_s$ is the velocity of sound and $q$ the absolute value of the momentum wave vector. The matrix element for electron-phonon coupling with this mode is obtained by deformation potential theory and reads \cite{Brouwer2014,brouwer2017,Kaiser2000}
\begin{equation}
\left\vert M^\text{LA}_{e-ph}(q)\right\vert ^2=\frac{\hbar C^2 q^2}{2\rho\omega_\text{LA}(q) (1+\left(q_0/q\right)^2)}\enspace ,
\end{equation}
where $C$ is a constant deformation potential and $\rho$ is the mass density.
Longitudinal optical (LO) phonons are approximated by the
Einstein model where the phonon frequency is constant, $\omega_{\text{LO}_{1,2}}(q)=\omega_{\text{LO}_{1,2}}$. 
For the interaction of electrons with these modes, we apply a polar matrix element, 
which reads~\cite{Brouwer2014,brouwer2017,Kaiser2000}
\begin{equation}\label{phmelo}
 \left\vert M^\text{LO}_{e-ph}(q)\right\vert ^2=\frac{e^2}{\varepsilon_0}\frac{\hbar\omega_{\text{LO}_{1,2}}}{2 q^2(1+\left(q_0/q\right)^2)}\left(\frac{1}{\varepsilon_\infty}-\frac{1}{\varepsilon_\text{r}}\right) \enspace ,
\end{equation}
where $\varepsilon_r$ is the relative permittivity and $\varepsilon_\infty$ the optical permittivity.

When we analyze non-equilibrium distributions it can be useful to define the quasi-temperature and chemical potential which allow us to compare the non-equilibrium distribution to an equilibrium distribution of the same particle- and energy density. They are found by first calculating the particle density
\begin{equation}
n_{\text{CB,VB}} = \int_{\text{CB,VB}} dE \, \mathrm{DOS}(E) f(E)
\label{eq:density}
\end{equation}
and the energy density
\begin{equation}
u_{\text{CB,VB}} = \int_{\text{CB,VB}} dE \, \mathrm{DOS}(E) f(E) E
\label{eq:energy}
\end{equation}
of the non-equilibrium distribution $f(E)$ in either the conduction or valence band, where $\mathrm{DOS}(E)$ is the density of states. Then, a root-finding algorithm is used to find the temperature $T$ and chemical potential $\mu$ of a Fermi distribution $f_{\text{Fermi}}(E,T,\mu)$ which yields the same particle and energy density in the considered band.

\section{Results}
\label{sec:results}

As in Refs.~\cite{Brouwer2014,brouwer2017}, we consider for the electrons two parabolic bands, namely an initially filled valence band with an effective mass of $- 3.51 \, m_e$ and an initially empty conduction band with an effective mass of $m_e$. 
The extrema of both bands are separated by a band gap of $9\usk\electronvolt$. The Fermi energy is located in the center of the band gap. 
Initially, all electrons are described by a single Fermi distribution of $T_e = 300\usk\kelvin$. 
The longitudinal acoustic phonon mode is considered with a sound velocity of $c_s = \unit{5935}{\metre\per\second}$, whereas the optical modes are considered with the Einstein model at frequencies of $\hbar \omega_1 = \unit{63}{\milli\electronvolt}$ and $\hbar \omega_2 = \unit{153}{\milli\electronvolt}$. We assume that all phonon modes extend over the whole spherical Brillouin zone with radius $q_B = 2.35 \cdot 10^{10}\usk\reciprocal\metre$.

To solve the 
system of coupled integro-differential equations given by Eqs.~\eqref{eqn:isobm} and \eqref{eqn:phbm}, the electronic energies are discretized by 3000 points and the phonon wavevectors by 75 points, respectively. 
To evaluate the electron--electron collision integrals,
we apply the MISER Monte Carlo integration algorithm~\cite{gsl}. All other collision integrals are solved using the trapezoidal rule.
The propagation in time is done by a 
forward Euler method with an adaptive time step.

In the following, we will present the electron distribution function $f$ in the so-called $\Phi$ representation
\cite{Kaiser2000,Rethfeld2002}, 
which is related to the distribution function through a logarithm as
\begin{equation}
\Phi(E) = -\log\!\left(\frac{1}{f(E)}-1\right)\,. \label{eqn:phi}
\end{equation}
In comparison to a simple logarithm, 
Eq.~(\ref{eqn:phi}) has the advantage that a Fermi distribution is represented by a linear function, also including the electrons in the valence band. It has a slope inversely proportional to the temperature of the system. 
In the case of a non-equilibrium distribution, the $\Phi$-representation deviates from a straight line,
allowing an easy and direct visual identification of deviations 
from equilibrium in both the conduction and the valence bands.

\subsection{Electron relaxation after laser irradiation}
\label{sec:electronrelax-wlaser-noeph}

In a first step, we analyze the behavior of the conduction and valence band electron distribution functions in our two-band model dielectric under laser irradiation while neglecting the phonon system.
To that end, we set the electron--phonon collision term in equation~\eqref{eqn:isobm} to zero.
Note that the e-ph-pt collision term is still considered, allowing for a realistic laser energy absorption.

As laser parameters, we chose a $\unit{150}{\femto\second}$ FWHM Gaussian laser pulse with a wavelength of $\unit{400}{\nano\meter}$ and a peak intensity of $1.33\cdot 10^{18}\usk\watt\per\meter\squared$.

\begin{figure}
	\centering
	\includegraphics[width=0.5\textwidth]{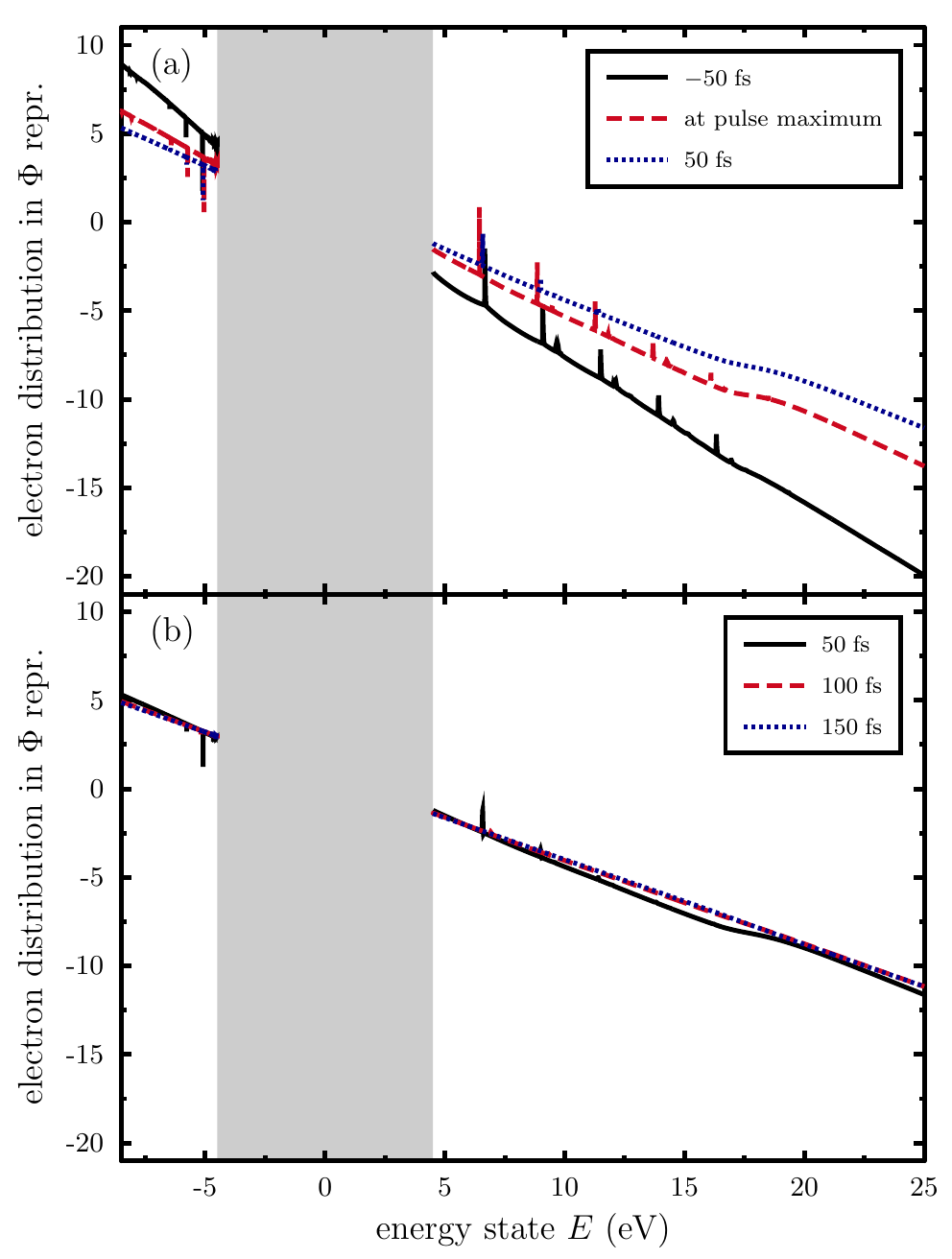}
	\caption{\label{fig:noephnoneqdist_all}Electron distribution in $\Phi$-representation \eqref{eqn:phi}, (a) during excitation of a model dielectric with a $150\usk\femto\second$ FWHM laser pulse and (b) shortly thereafter. The gray shadowed area indicates the band gap. Energies above the band gap belong to the conduction band, energies below the band gap to the valence band. For this simulation, the coupling to the phonon system was neglected. In this representation, straight lines correspond to a Fermi distribution. }
\end{figure}

Figure\,\ref{fig:noephnoneqdist_all} shows the electron distribution in $\Phi$-re\-pre\-sen\-tation during and shortly after laser irradiation. The gray shaded area indicates the band gap. The energies above the band gap belong to the conduction band, while the energies below the band gap belong to the valence band.
The distribution at $-50\usk\femto\second$, i.e.  before the maximum laser intensity, shows peaks in
the conduction band, arising from a combination of multi-photon ionization, including above-threshold ionization, and subsequent intraband single-photon absorption \cite{Brouwer2014}.
In the valence band, corresponding dips appear, indicating a lack of electrons, or the presence of holes, respectively.
At the temporal maximum of the laser pulse
the distribution has risen in the conduction band and fallen in the valence band, as more electrons have been promoted from one band to the other.
The peak-and-dip structure still exists, but in addition a bump is visible at higher energies in the conduction band.
This bump is caused by Auger recombination, which in this case exceeds impact ionization~\cite{Medvedev2011}.

The electron distribution $50 \usk\femto\second$ after the maximum of the Gaussian laser pulse is shown in both parts of Fig.~\ref{fig:noephnoneqdist_all}.
The peaks and dips caused by multi-photon ionization as well as the bump at higher energies in the conduction band caused by Auger recombination are still present. 
In contrast to that, the electron distribution is nearly equilibrated after $100 \usk\femto\second$, as represented by the nearly straight $\Phi$-functions, see Eq.~\eqref{eqn:phi}.

\begin{figure}
	\centering
	\includegraphics[width=0.5\textwidth]{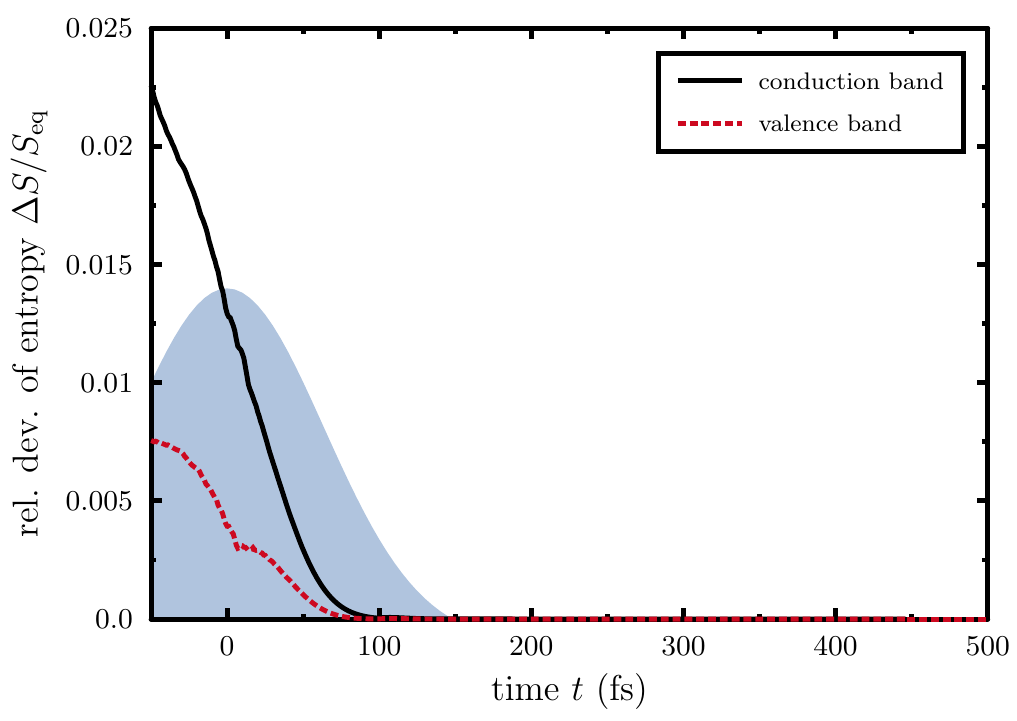}
	\caption{\label{fig:noephentropy}Relative deviation of the configuration entropy of the non-equilibrium distribution from the entropy of a Fermi distribution with the same particle and energy content within the conduction and valence band, respectively.}
\end{figure}

To further study the equilibration of the electron systems, we introduce the configuration entropy~\cite{landau1980lifshitz}
\begin{align}
S = -k_B \int& dE \, \mathrm{DOS}(E) \, \Big [ f(E) \log(f(E)) \nonumber \\
&+ (1 - f(E)) \log(1-f(E)) \Big ] \, ,
\label{eq:entropy}
\end{align}
where $k_B$ is Boltzmann's constant.
By comparing the configuration entropy of the conduction and valence band to the entropy of the corresponding Fermi distribution, the progress of intraband thermalization can be investigated. Figure \ref{fig:noephentropy} depicts the relative deviation of the configuration entropy compared to the equilibrium entropy. 
Both the deviations of the conduction band and valence band are falling shortly before the pulse maximum and thereafter, and an intraband equilibrium is reached about $100 \usk\femto\second$ after the pulse maximum for both bands. The initial deviation is higher for the conduction band due to its lower effective mass compared to the valence band. 

\begin{figure}
	\centering
	\includegraphics[width=0.5\textwidth]{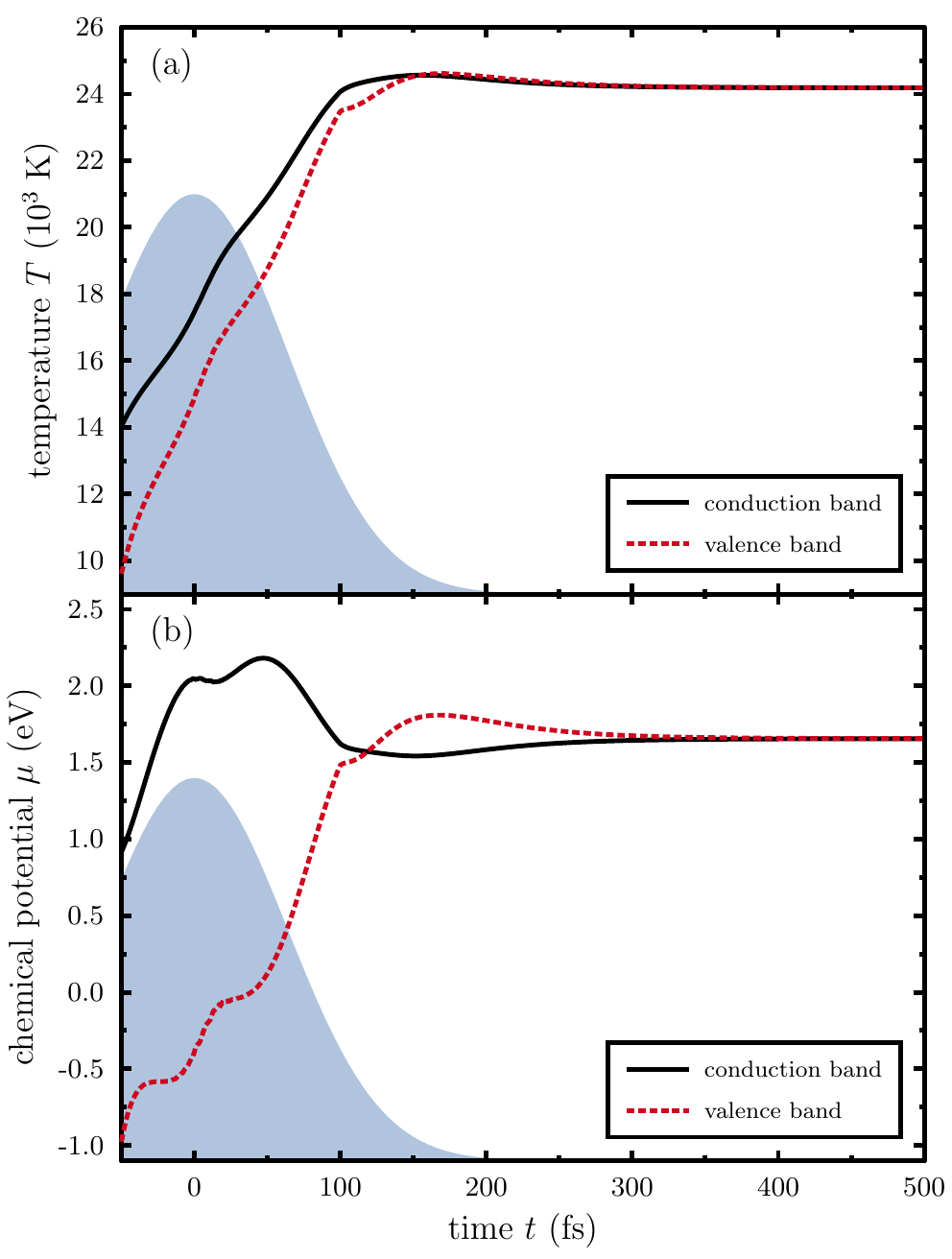}
	\caption{\label{fig:lastemp_mu1}(a) Electron (quasi-) temperature and (b) chemical potential of valence and conduction band electrons during and after laser excitation, neglecting the coupling to the phonon system. The shading indicates the temporal shape of the laser pulse.}
\end{figure}

Next, we want to investigate the interband relaxation by comparing the temporal evolution of the quasi-temperatures and chemical potentials of both bands.
The temporal evolution of the quasi-temperatures is shown in Fig.~\ref{fig:lastemp_mu1}~(a).
Both the conduction and valence band quasi-temperatures are rising during the laser irradiation and reach the same level approximately $100\usk\femto\second$ after the laser pulse maximum.
Analogously, the temporal evolution of the chemical potentials is depicted in Fig.~\ref{fig:lastemp_mu1}~(b).
While both the  conduction and valence band chemical potential rise initially, their difference increases approximately up to the pulse maximum. For the chosen laser parameters, the laser irradiation excited more electrons to the conduction band than the absorbed energy would support in equilibrium. This results in a difference between the (quasi-) chemical potentials of the conduction and valence bands. The non-equilibrium between the chemical potentials is counteracted by Auger recombination, which decreases the conduction band electron density and increases the valence band electron density. 

During the falling tail of the laser pulse, linear intraband absorption by e-ph-pt collisions becomes more important  than multi-photon absorption at lower intensities. The e-ph-pt collisions increase the energy of the conduction and valence bands, so that again more free electrons and holes are supported in equilibrium. This leads to a faster equilibration of the chemical potentials by Auger recombination alone. About $100\usk\femto\second$ after the pulse maximum, the chemical potentials reach the same level and then slightly overshoot due to the heating by e-ph-pt collisions, so that the chemical potential of the valence band becomes larger than the chemical potential of the conduction band. Then impact ionization exceeds over Auger recombination so that electrons are transferred from the valence band to the conduction band. After about $300 \usk\femto\second$ the chemical potentials finally equilibrate.

In this section we investigated the intra- and interband thermalization of the electron bands during and shortly after laser irradiation, neglecting electron-phonon coupling. In those circumstances, we see fast intraband thermalization and equilibration of the quasi-temperatures between the bands on a timescale of about $100 \usk\femto\second$. The equilibration of the chemical potentials takes slightly longer and is influenced by an interplay between nonlinear multiphoton ionization,  linear e-ph-pt collisions
and interband relaxation processes. In the next section, we separate these excitation and relaxation processes.

\subsection{Without Laser}
\label{sec:nolaser}
In this section, we replace the initial laser excitation by excited Fermi distributions in the conduction and valence bands as initial condition. This allows us to determine the intrinsic relaxation times for the system under study.
First we discuss the qualitative behavior of the dynamics of temperatures and chemical potentials during interband relaxation. Later in this section we will investigate the dependence of the relaxation times on different parameters.

\begin{figure}
	\centering
	\includegraphics[width=0.5\textwidth]{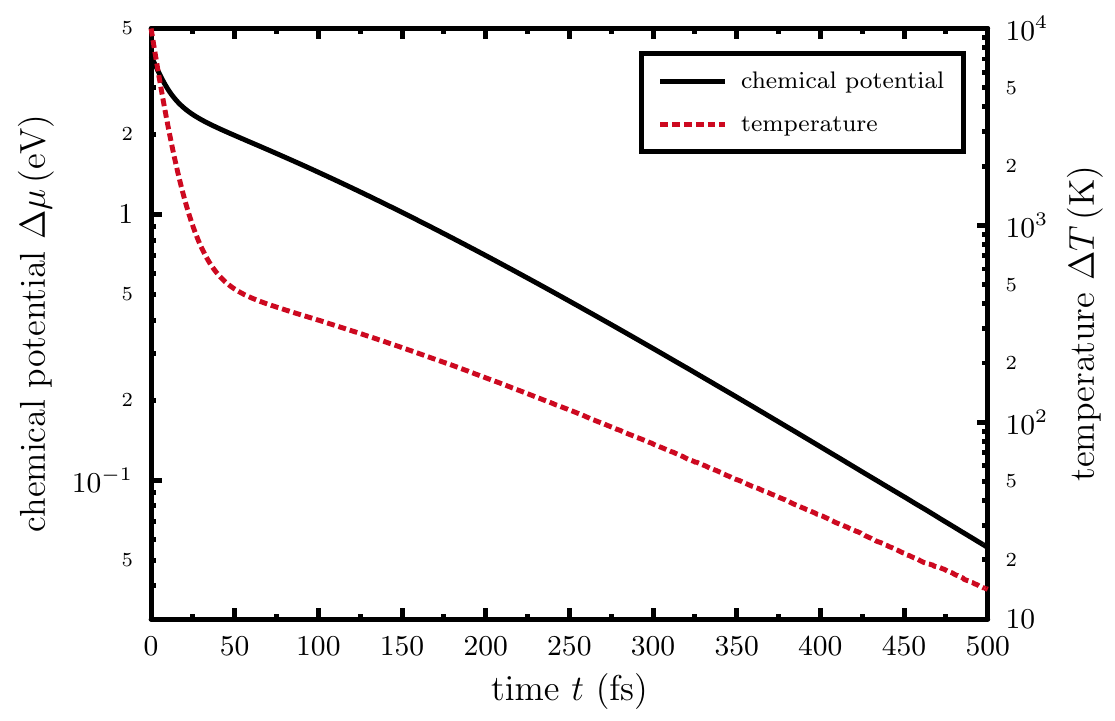}
	\caption{\label{fig:deltat}Temporal evolution of the electron temperature difference and chemical potential difference, during interband relaxation. Initial conditions: $T_c=20000\usk\kelvin$ and $T_v=10000\usk\kelvin$; fraction of $1\%$ of valence band electrons excited to the conduction band.}
\end{figure}

Figure \ref{fig:deltat} shows the temporal evolution of the difference between the chemical potentials and temperatures, respectively, of the conduction and valence bands. Initially, $1 \%$ of the valence band electrons are excited into the conduction band and the conduction band temperature is set to $T_e=20000\usk\kelvin$ and the valence band temperature to $T_v=10000\usk\kelvin$. For those parameters, for both the difference in temperature and chemical potential, a two-timescale behavior is observed: At first a faster relaxation on the timescale of less than $10\usk\femto\second$ occurs, followed by a slower relaxation on the timescale of about $150\usk\femto\second$. The first faster relaxation can be attributed to energy-exchanging interband electron-hole collisions, which drive the quasi-temperatures of the two bands into equilibrium. The following slower relaxation is governed by particle exchanging processes, that is Auger recombination and impact ionization. While the latter processes contribute to the equilibration of the chemical potentials of valence and conduction band, they also have a delaying effect on the temperature relaxation due to the conversion between kinetic and potential energy.

\begin{figure}
	\centering
	\includegraphics[width=0.5\textwidth]{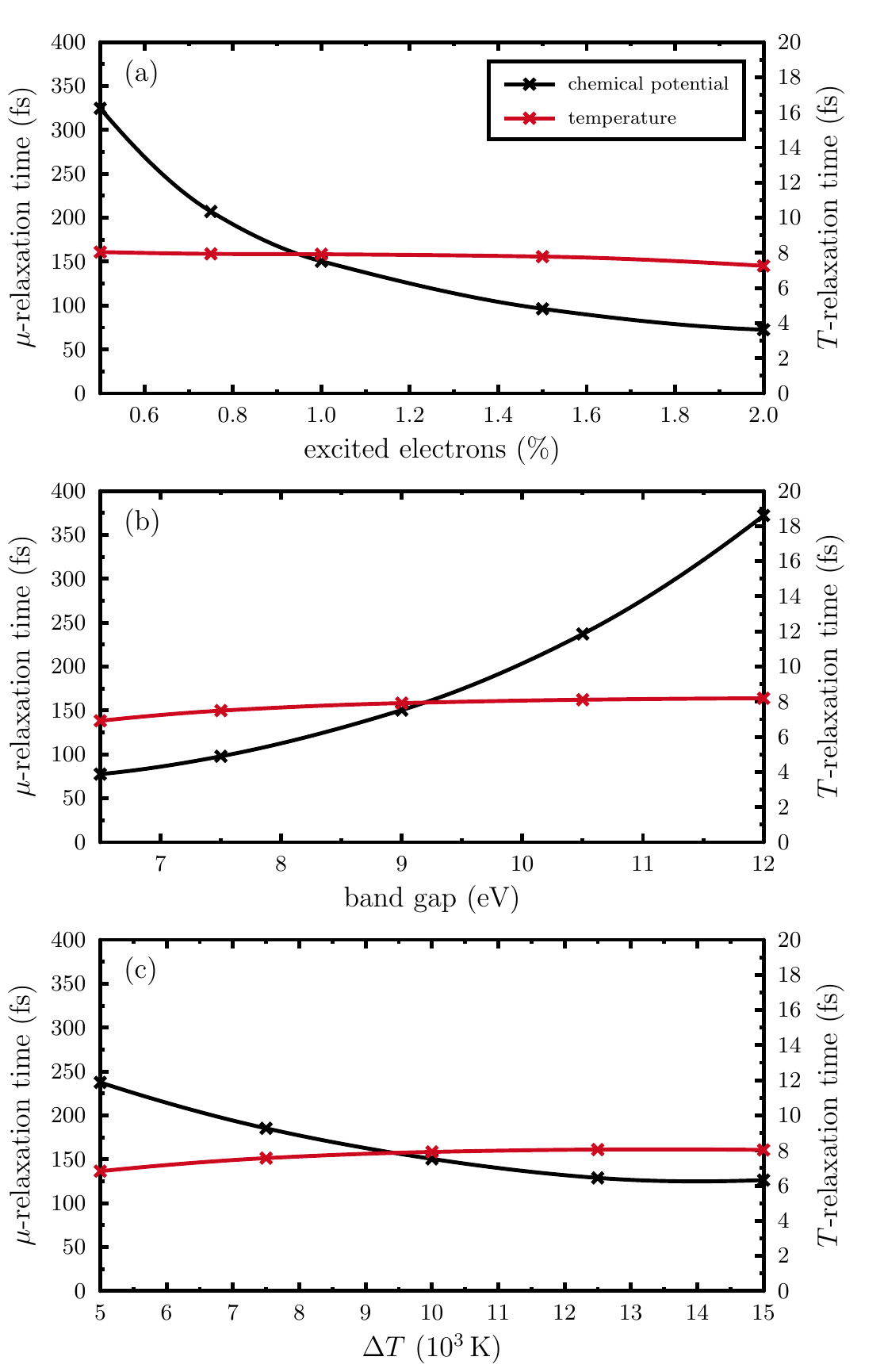}
	\caption{\label{fig:taumu_all}Chemical potential relaxation time in dependence on (a) initial percentage of excited electrons, (b) band gap and (c) initial temperature difference between conduction and valence band. If the corresponding parameter is not varied, the initial temperatures are $T_e=20000\usk\kelvin$ and $T_v=10000\usk\kelvin$, initially $1\%$ of all electrons are excited into the conduction band and the band gap is set to $9\usk\electronvolt$.}
\end{figure}

The fast and the slow relaxation underlying the behavior in Fig.~\ref{fig:deltat} can be described with each a relaxation time, which we will denote as $\tau_T$ and $\tau_\mu$. Here, $\tau_T$ is the fast relaxation time which we attributed to energy-exchanging electron-hole collisions driven by a temperature difference and $\tau_\mu$ is the slower relaxation time which we attributed to particle exchange by Auger recombination and impact ionization driven by a difference in chemical potentials.
We fit a double exponential function
\begin{equation}\label{eq fit}
\Delta X(t) = A e^{-t/\tau_T} + B e^{-t/\tau_\mu},
\end{equation}
to the temporal evolution of the difference of the temperatures 
or the chemical potentials
, respectively, for the first $200\usk\femto\second$, in order to obtain the fit parameters $A$, $B$, $\tau_T$ and $\tau_\mu$.  
The resulting characteristic times $\tau_T$ and $\tau_\mu$ are similar for a fit to either $X = \mu$ or $X = T$.
Therefore, we will show in the following the $T$-relaxation time $\tau_T$ obtained by a fit to the temperature difference $\Delta T$,  and the $\mu$-relaxation time $\tau_\mu$ obtained by a fit to the chemical potential difference $\Delta \mu$. For the following results, we chose parameter ranges for the starting conditions, for which the temporal evolution clearly yields a double exponential behavior. 
We determine this by an adjusted R-squared value above $0.99$ of the fit \eqref{eq fit}. At for example higher amounts of initially excited electrons for the same initial temperatures, the equilibration of the chemical potentials by Auger recombination can have a larger effect on the conduction band temperature than the equilibration of the temperatures due to electron-hole collisions. In this case, the temperature difference increases initially and the two timescales can not be separated. 

Figure \ref{fig:taumu_all} shows the relaxation time for the chemical potentials and temperatures in dependence on different parameters.
For all figures, the initial valence band temperature is $T_v = 10000\usk\kelvin$. The initial conduction band temperature is $T_c = T_v + \Delta T$. If the parameter is not the varied parameter, $\Delta T = 10000\usk\kelvin$, the band gap equals $9\usk\electronvolt$ and $1\%$ of the valence band electrons are initially excited to the conduction band.
The crosses represent the calculated values and the solid curves are fits that are shown to guide the eyes.

In Fig.~\ref{fig:taumu_all} (a) the percentage of initially excited electrons is varied. The $\mu$--relaxation time $\tau_\mu$ shows a significant dependence on the percentage of excited electrons. While an interband relaxation time of $\sim 300\usk\femto\second$ is observed for $0.5 \%$ initially excited electrons, the relaxation time is decreased to $\sim 70\usk\femto\second$ for $2 \%$.
The $T$-relaxation time $\tau_T$ only slightly varies around $8\usk\femto\second$.

The relaxation times in dependence on the band gap are depicted in Fig.~\ref{fig:taumu_all} (b). A significant increase in the $\mu$--relaxation time is observed with increasing band gap, while the initial percentage of excited electrons and the temperatures are kept constant. It should be noted that the same laser pulse would excite more electrons to the conduction band for a material with a smaller band gap than for a material with a larger band gap. The $T$--relaxation time shows no significant dependence on the band gap energy.

Figure \ref{fig:taumu_all} (c) shows the dependence of the relaxation times on the difference of the initial temperatures. The initial valence band temperature is thereby kept at $10000\usk\kelvin$ and the conduction band temperature is varied. Therefore, the energy content increases with $\Delta T$. A decrease of the $\mu$--relaxation time with increasing temperature difference or increasing energy content can be observed. The $T$--relaxation time only varies slightly.

Our results show a significant influence of the excitation strength, as simulated by starting with varying percentages of excited electrons and temperature differences on the $\mu$--relaxation time. Generally, we observe a larger $\mu$--relaxation time for stronger excitations. Also, the $\mu$--relaxation time is larger for larger band gap energies. The $T$--relaxation time only varies slightly for all considered parameters. 
In all cases, the temperature relaxation is considerably faster than the equilibration of the chemical potentials, so that each band first establishes its own Fermi distribution at equal temperatures in the conduction and valence bands, before density-changing scattering processes lead to a common Fermi distribution across both bands. 

\subsection{Influence of electron-phonon coupling}
\label{sec:electronrelax-wlaser-weph}

In this section we study the electron relaxation after laser irradiation, when in contrast to section \ref{sec:electronrelax-wlaser-noeph} electron-phonon coupling is considered.

\begin{figure}
	\centering
	\includegraphics[width=0.5\textwidth]{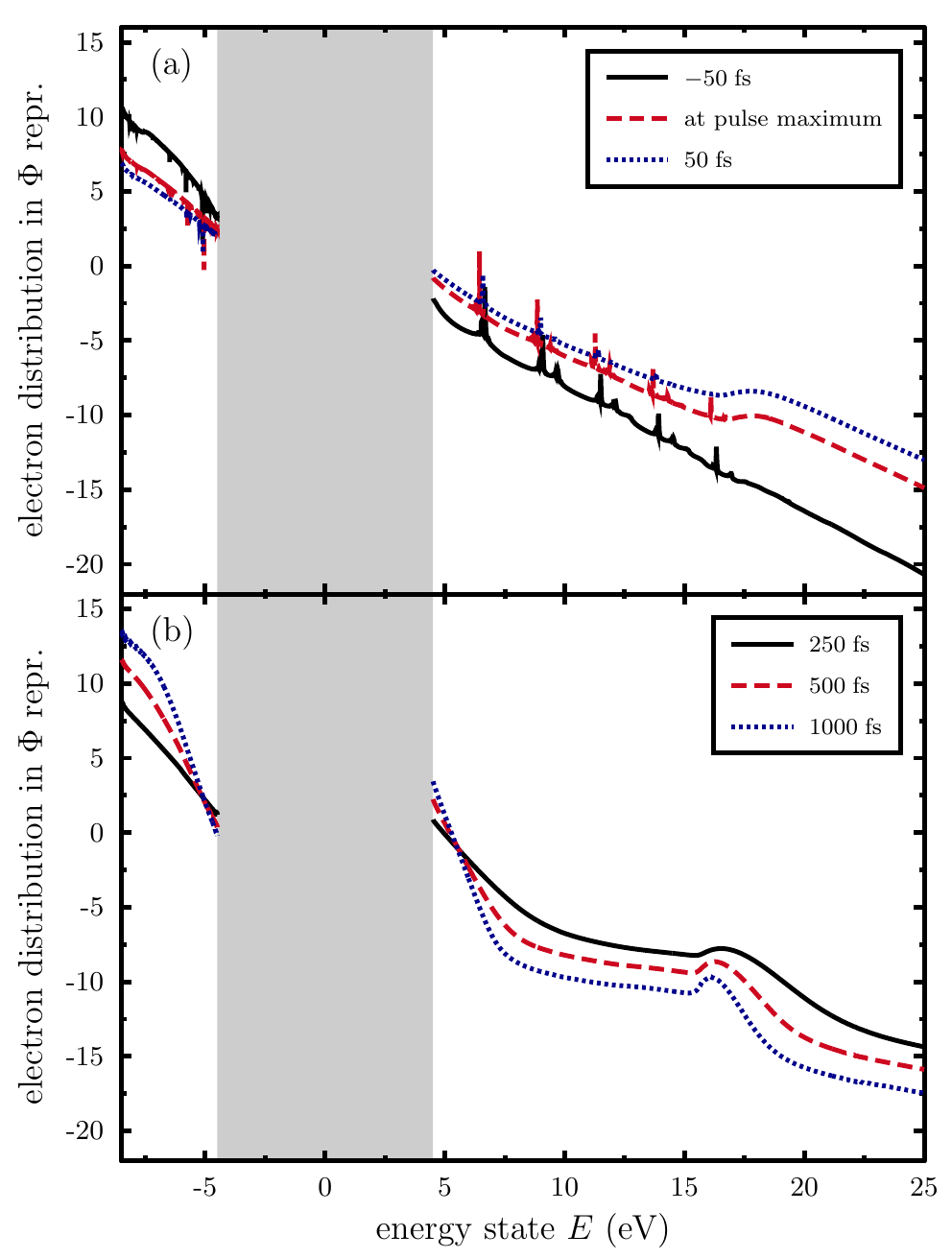}
	\caption{\label{fig:noneqdist_all}Electron distribution in $\Phi$ representation \eqref{eqn:phi}, (a) during excitation of a model dielectric with a $150\usk\femto\second$ FWHM laser pulse and (b) shortly thereafter. Note that straight lines correspond to a Fermi distribution in this representation. In contrast to Fig.~\ref{fig:noephnoneqdist_all}, the coupling to the phonon system was included here.}
\end{figure}

Figure \ref{fig:noneqdist_all} shows the electron distribution at different times (a) during laser excitation and (b) thereafter for the full simulation including electron-phonon coupling.
Compared to the electronic distributions during laser excitation in Fig.~\ref{fig:noephnoneqdist_all}~(a), where electron-phonon coupling was not taken into account, the same features can be observed. However, the bump at higher energies caused by Auger recombination is more noticeable for the curves shown in Fig.~\ref{fig:noneqdist_all}~(a).

The electron distributions for later times after laser excitation are presented in Fig.~\ref{fig:noneqdist_all}~(b). In contrast to the fast equilibration shown in figure \ref{fig:noephnoneqdist_all}~(b) without electron-phonon coupling, the electron distribution still deviates noticeably from a Fermi distribution as represented by a straight line after $1000\usk\femto\second$ for the full simulation. Specifically, the Auger bump at high energies in the conduction band persists for a long time.
Note that such Auger bump has been observed also in XUV-excited aluminum \cite{Medvedev2011}.

\begin{figure}
	\centering
	\includegraphics[width=0.5\textwidth]{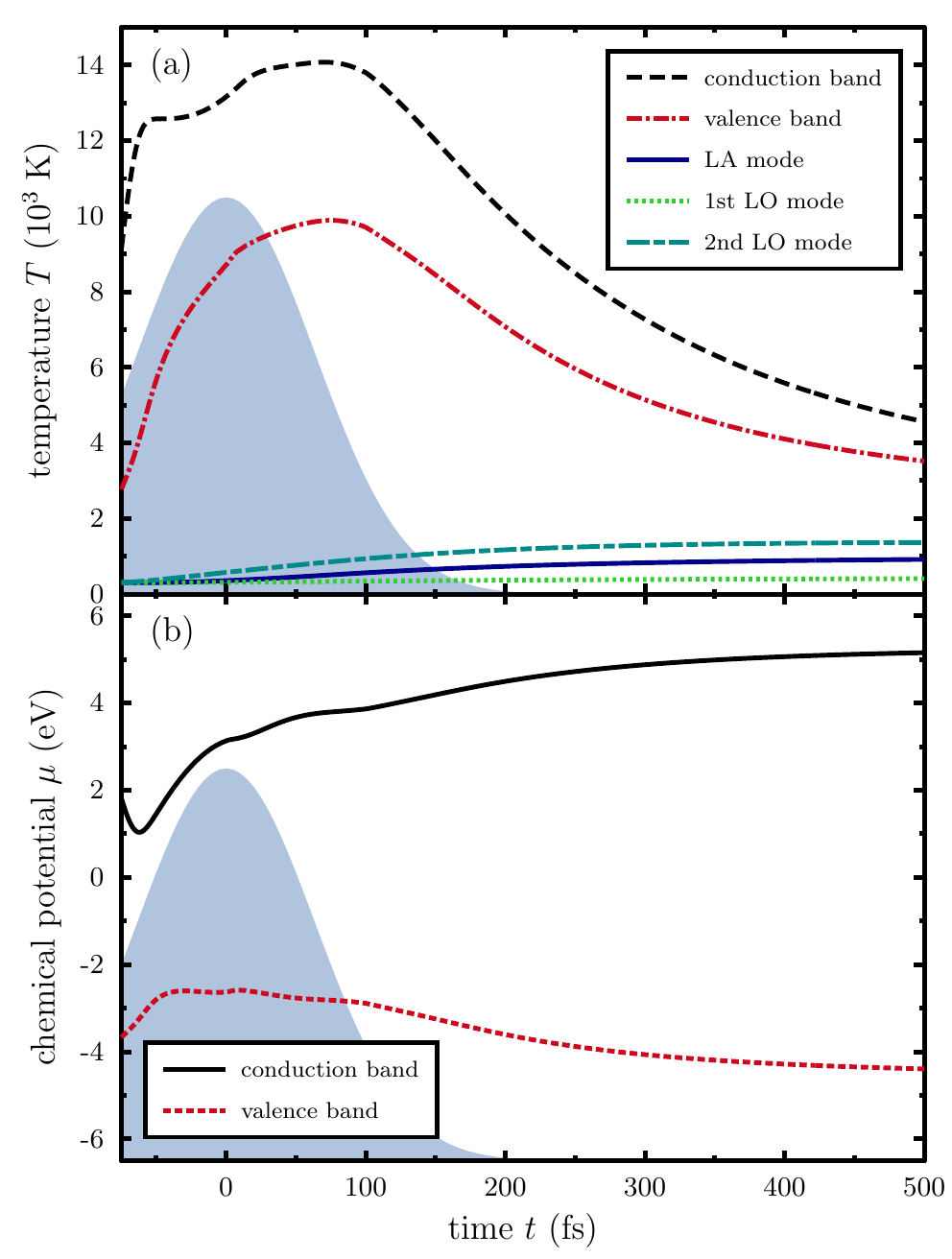}
	\caption{\label{fig:mu150_t150_all}(a) Electron and phonon (quasi-) temperatures and (b) chemical potentials during and after laser excitation. In contrast to Fig.~\ref{fig:lastemp_mu1}, the coupling to the phonon system was included here.}
\end{figure}

Next, we want to investigate interband thermalization by considering the (quasi-) temperatures and chemical potentials of the different systems. 
Figure \ref{fig:mu150_t150_all}~(a) shows the temporal evolution of the temperatures of the conduction and valence band electrons as well as the three considered phonon modes.
The electron temperatures first rise during laser excitation and reach a maximum about $100\usk\femto\second$ after the pulse maximum. Thereafter, the electron temperatures decrease and phonon temperatures increase due to electron-phonon coupling. Compared to figure \ref{fig:lastemp_mu1}~(a) where electron-phonon coupling was neglected, the temperature difference between conduction and valence band persists for a much longer time.

The temporal evolution of the chemical potentials of conduction and valence band is shown in figure \ref{fig:mu150_t150_all}~(b). In contrast to the case of neglecting electron-phonon coupling as depicted in figure \ref{fig:lastemp_mu1}~(b), the difference between the chemical potentials increases during and after laser excitation.
This can be explained by the thermalization of the electrons with the phonon system, which causes the energy content of the electrons to decrease after laser irradiation. As the energy content decreases, less excited electrons are supported by the corresponding equilibrium distribution, which counteracts the equilibration of the chemical potentials by Auger recombination. Additionally, the Auger recombination rate decreases with decreasing energy in the electron system such that the equilibration of chemical potentials is further slowed down.

Our results show that both intra- and interband thermalization are  considerably delayed by electron-phonon coupling. The chemical potentials of the conduction and valence bands are even driven further out of interband equilibrium shortly after laser excitation, leading to a situation far from equilibrium on the timescale of electron--phonon equilibration.
Since the electron--phonon coupling itself depends on the density of excited electrons as well as directly on the electronic energy distribution \cite{brouwer2017}, the complex full equilibration process needs further studies.

\section{Summary and Conclusions}
\label{sec:summary}
In this work, we studied the relaxation of the conduction band, valence band and phonon system in a dielectric after excitation with an intense, ultrafast laser pulse.  
We have calculated the time evolution of the laser-excited electron distribution functions in the valence and conduction bands in a large bandgap material. We observe a peak-like structure of the energy distribution in the conduction and valence bands, and a fast relaxation to an equilibrium situation, which is a joint Fermi distribution across both bands. 
Both bands equilibrate to a thermal distribution already during the time of laser irradiation. Both temperatures equilibrate to a joint temperature within about ten femtoseconds, while an occupational nonequilibrium, characterized by two different chemical potentials, may last up to a few hundreds of femtoseconds.  

We have quantified two distinct characteristic relaxation times, $\tau_T$ and $\tau_\mu$, which we then studied in dependence on material- and laser-parameters for well-defined initial situations. The relaxation time of the chemical potential turned out to be more sensitive to the studied parameters than the relaxation time of the temperature difference. 

We then included electron--phonon coupling in our calculations, which strongly perturbs the separation of timescales.  
Instead of a fast equilibration towards a Fermi distribution at elevated temperature, the simultaneous cooling of the electron gas leads to a continuous overpopulation of the conduction band. This overpopulation boosts Auger recombination, which reduces the total number of excited electrons but 
feeds electrons at high energy states into the conduction band. 
The resulting energy distribution in both considered bands differs strongly from a Fermi distribution.

We therefore conclude that an athermal electron distribution can be present in the material far into the picosecond timescales.

\begin{acknowledgments}
The authors 
gratefully acknowledge funding by the Deutsche
Forschungsgemeinschaft (DFG, German Research Foundation) - TRR 173 - 268565370 Spin + X: spin in
its collective environment (Project A08), 
as well as computing resources 
on the Elwetritsch high performance computing cluster granted by the AHRP under project TUK-STREMON.
We thank Sebastian T.~Weber and Dirk O.~Gericke for fruitful discussions.
\end{acknowledgments}

% Create the reference section using BibTeX:
\bibliography{all}

%apsrev4-2.bst 2019-01-14 (MD) hand-edited version of apsrev4-1.bst
%Control: key (0)
%Control: author (8) initials jnrlst
%Control: editor formatted (1) identically to author
%Control: production of article title (0) allowed
%Control: page (0) single
%Control: year (1) truncated
%Control: production of eprint (0) enabled
\begin{thebibliography}{36}%
\makeatletter
\providecommand \@ifxundefined [1]{%
 \@ifx{#1\undefined}
}%
\providecommand \@ifnum [1]{%
 \ifnum #1\expandafter \@firstoftwo
 \else \expandafter \@secondoftwo
 \fi
}%
\providecommand \@ifx [1]{%
 \ifx #1\expandafter \@firstoftwo
 \else \expandafter \@secondoftwo
 \fi
}%
\providecommand \natexlab [1]{#1}%
\providecommand \enquote  [1]{``#1''}%
\providecommand \bibnamefont  [1]{#1}%
\providecommand \bibfnamefont [1]{#1}%
\providecommand \citenamefont [1]{#1}%
\providecommand \href@noop [0]{\@secondoftwo}%
\providecommand \href [0]{\begingroup \@sanitize@url \@href}%
\providecommand \@href[1]{\@@startlink{#1}\@@href}%
\providecommand \@@href[1]{\endgroup#1\@@endlink}%
\providecommand \@sanitize@url [0]{\catcode `\\12\catcode `\$12\catcode
  `\&12\catcode `\#12\catcode `\^12\catcode `\_12\catcode `\%12\relax}%
\providecommand \@@startlink[1]{}%
\providecommand \@@endlink[0]{}%
\providecommand \url  [0]{\begingroup\@sanitize@url \@url }%
\providecommand \@url [1]{\endgroup\@href {#1}{\urlprefix }}%
\providecommand \urlprefix  [0]{URL }%
\providecommand \Eprint [0]{\href }%
\providecommand \doibase [0]{https://doi.org/}%
\providecommand \selectlanguage [0]{\@gobble}%
\providecommand \bibinfo  [0]{\@secondoftwo}%
\providecommand \bibfield  [0]{\@secondoftwo}%
\providecommand \translation [1]{[#1]}%
\providecommand \BibitemOpen [0]{}%
\providecommand \bibitemStop [0]{}%
\providecommand \bibitemNoStop [0]{.\EOS\space}%
\providecommand \EOS [0]{\spacefactor3000\relax}%
\providecommand \BibitemShut  [1]{\csname bibitem#1\endcsname}%
\let\auto@bib@innerbib\@empty
%</preamble>
\bibitem [{\citenamefont {B\"{a}uerle}(2011)}]{BaeuerleBuch11}%
  \BibitemOpen
  \bibfield  {author} {\bibinfo {author} {\bibfnamefont {D.}~\bibnamefont
  {B\"{a}uerle}},\ }\href@noop {} {\emph {\bibinfo {title} {Laser Processing
  and Chemistry}}}\ (\bibinfo  {publisher} {Springer Verlag},\ \bibinfo
  {address} {Berlin, Heidelberg},\ \bibinfo {year} {2011})\BibitemShut
  {NoStop}%
\bibitem [{\citenamefont {Vogel}\ \emph {et~al.}(2005)\citenamefont {Vogel},
  \citenamefont {Noack}, \citenamefont {H\"{u}ttman},\ and\ \citenamefont
  {Paltauf}}]{Vogel2005}%
  \BibitemOpen
  \bibfield  {author} {\bibinfo {author} {\bibfnamefont {A.}~\bibnamefont
  {Vogel}}, \bibinfo {author} {\bibfnamefont {J.}~\bibnamefont {Noack}},
  \bibinfo {author} {\bibfnamefont {G.}~\bibnamefont {H\"{u}ttman}},\ and\
  \bibinfo {author} {\bibfnamefont {G.}~\bibnamefont {Paltauf}},\ }\bibfield
  {title} {\bibinfo {title} {{Mechanisms of femtosecond laser nanosurgery of
  cells and tissues}},\ }\href {https://doi.org/10.1007/s00340-005-2036-6}
  {\bibfield  {journal} {\bibinfo  {journal} {{Appl. Phys. B}}\ }\textbf
  {\bibinfo {volume} {81}},\ \bibinfo {pages} {1015} (\bibinfo {year}
  {2005})}\BibitemShut {NoStop}%
\bibitem [{\citenamefont {Mangin}\ \emph {et~al.}(2014)\citenamefont {Mangin},
  \citenamefont {Gottwald}, \citenamefont {Lambert}, \citenamefont {Steil},
  \citenamefont {Uhl{\'\i}{\v{r}}}, \citenamefont {Pang}, \citenamefont {Hehn},
  \citenamefont {Alebrand}, \citenamefont {Cinchetti}, \citenamefont
  {Malinowski} \emph {et~al.}}]{Mangin2014NatureMaterials}%
  \BibitemOpen
  \bibfield  {author} {\bibinfo {author} {\bibfnamefont {S.}~\bibnamefont
  {Mangin}}, \bibinfo {author} {\bibfnamefont {M.}~\bibnamefont {Gottwald}},
  \bibinfo {author} {\bibfnamefont {C.}~\bibnamefont {Lambert}}, \bibinfo
  {author} {\bibfnamefont {D.}~\bibnamefont {Steil}}, \bibinfo {author}
  {\bibfnamefont {V.}~\bibnamefont {Uhl{\'\i}{\v{r}}}}, \bibinfo {author}
  {\bibfnamefont {L.}~\bibnamefont {Pang}}, \bibinfo {author} {\bibfnamefont
  {M.}~\bibnamefont {Hehn}}, \bibinfo {author} {\bibfnamefont {S.}~\bibnamefont
  {Alebrand}}, \bibinfo {author} {\bibfnamefont {M.}~\bibnamefont {Cinchetti}},
  \bibinfo {author} {\bibfnamefont {G.}~\bibnamefont {Malinowski}}, \emph
  {et~al.},\ }\bibfield  {title} {\bibinfo {title} {Engineered materials for
  all-optical helicity-dependent magnetic switching},\ }\href@noop {}
  {\bibfield  {journal} {\bibinfo  {journal} {Nature materials}\ }\textbf
  {\bibinfo {volume} {13}},\ \bibinfo {pages} {286} (\bibinfo {year}
  {2014})}\BibitemShut {NoStop}%
\bibitem [{\citenamefont {Keldysh}(1965)}]{Keldysh1965b}%
  \BibitemOpen
  \bibfield  {author} {\bibinfo {author} {\bibfnamefont {L.~V.}\ \bibnamefont
  {Keldysh}},\ }\bibfield  {title} {\bibinfo {title} {{Ionization in the field
  of a strong electromagnetic wave}},\ }\href
  {http://jetp.ac.ru/cgi-bin/dn/e_020_05_1307.pdf} {\bibfield  {journal}
  {\bibinfo  {journal} {Sov. Phys. JETP}\ }\textbf {\bibinfo {volume} {20}},\
  \bibinfo {pages} {1307} (\bibinfo {year} {1965})}\BibitemShut {NoStop}%
\bibitem [{\citenamefont {Balling}\ and\ \citenamefont
  {Schou}(2013)}]{Balling2013}%
  \BibitemOpen
  \bibfield  {author} {\bibinfo {author} {\bibfnamefont {P.}~\bibnamefont
  {Balling}}\ and\ \bibinfo {author} {\bibfnamefont {J.}~\bibnamefont
  {Schou}},\ }\bibfield  {title} {\bibinfo {title} {Femtosecond-laser ablation
  dynamics of dielectrics: basics and applications for thin films},\ }\href
  {https://doi.org/10.1088/0034-4885/76/3/036502} {\bibfield  {journal}
  {\bibinfo  {journal} {Reports on progress in physics}\ }\textbf {\bibinfo
  {volume} {76}},\ \bibinfo {pages} {036502} (\bibinfo {year}
  {2013})}\BibitemShut {NoStop}%
\bibitem [{\citenamefont {Koopmans}\ \emph {et~al.}(2010)\citenamefont
  {Koopmans}, \citenamefont {Longa}, \citenamefont {Malinowski}, \citenamefont
  {Steiauf}, \citenamefont {F\"{a}hnle}, \citenamefont {Roth}, \citenamefont
  {Cinchetti},\ and\ \citenamefont {Aeschlimann}}]{Koopmans2010}%
  \BibitemOpen
  \bibfield  {author} {\bibinfo {author} {\bibfnamefont {B.}~\bibnamefont
  {Koopmans}}, \bibinfo {author} {\bibfnamefont {F.~D.}\ \bibnamefont {Longa}},
  \bibinfo {author} {\bibfnamefont {G.}~\bibnamefont {Malinowski}}, \bibinfo
  {author} {\bibfnamefont {D.}~\bibnamefont {Steiauf}}, \bibinfo {author}
  {\bibfnamefont {M.}~\bibnamefont {F\"{a}hnle}}, \bibinfo {author}
  {\bibfnamefont {T.}~\bibnamefont {Roth}}, \bibinfo {author} {\bibfnamefont
  {M.}~\bibnamefont {Cinchetti}},\ and\ \bibinfo {author} {\bibfnamefont
  {M.}~\bibnamefont {Aeschlimann}},\ }\bibfield  {title} {\bibinfo {title}
  {Explaining the paradoxical diversity of ultrafast laser-induced
  demagnetization},\ }\href {https://doi.org/10.1038/nmat2593} {\bibfield
  {journal} {\bibinfo  {journal} {Nature Materials}\ }\textbf {\bibinfo
  {volume} {9}},\ \bibinfo {pages} {259} (\bibinfo {year} {2010})}\BibitemShut
  {NoStop}%
\bibitem [{\citenamefont {Bauer}\ \emph {et~al.}(2015)\citenamefont {Bauer},
  \citenamefont {Marienfeld},\ and\ \citenamefont {Aeschlimann}}]{Bauer2015}%
  \BibitemOpen
  \bibfield  {author} {\bibinfo {author} {\bibfnamefont {M.}~\bibnamefont
  {Bauer}}, \bibinfo {author} {\bibfnamefont {A.}~\bibnamefont {Marienfeld}},\
  and\ \bibinfo {author} {\bibfnamefont {M.}~\bibnamefont {Aeschlimann}},\
  }\bibfield  {title} {\bibinfo {title} {Hot electron lifetimes in metals
  probed by time-resolved two-photon photoemission},\ }\href
  {https://doi.org/10.1016/j.progsurf.2015.05.001} {\bibfield  {journal}
  {\bibinfo  {journal} {Progress in Surface Science}\ }\textbf {\bibinfo
  {volume} {90}},\ \bibinfo {pages} {319 } (\bibinfo {year}
  {2015})}\BibitemShut {NoStop}%
\bibitem [{\citenamefont {Winkler}\ \emph {et~al.}(2017)\citenamefont
  {Winkler}, \citenamefont {Haar.Lillevang}, \citenamefont {Sarpe},
  \citenamefont {Zielinski}, \citenamefont {Götte}, \citenamefont {Balling},\
  and\ \citenamefont {Baumert}}]{Winkler2017}%
  \BibitemOpen
  \bibfield  {author} {\bibinfo {author} {\bibfnamefont {T.}~\bibnamefont
  {Winkler}}, \bibinfo {author} {\bibfnamefont {L.}~\bibnamefont
  {Haar.Lillevang}}, \bibinfo {author} {\bibfnamefont {C.}~\bibnamefont
  {Sarpe}}, \bibinfo {author} {\bibfnamefont {B.}~\bibnamefont {Zielinski}},
  \bibinfo {author} {\bibfnamefont {A.}~\bibnamefont {Götte}, \bibfnamefont
  {N.~Senftleben}}, \bibinfo {author} {\bibfnamefont {P.}~\bibnamefont
  {Balling}},\ and\ \bibinfo {author} {\bibfnamefont {T.}~\bibnamefont
  {Baumert}},\ }\bibfield  {title} {\bibinfo {title} {Laser amplification in
  excited dielectrics},\ }\bibfield  {journal} {\bibinfo  {journal} {nature
  physics}\ }\href {https://doi.org/10.1038/NPHYS4265} {10.1038/NPHYS4265}
  (\bibinfo {year} {2017})\BibitemShut {NoStop}%
\bibitem [{\citenamefont {Shugaev}\ \emph {et~al.}(2016)\citenamefont
  {Shugaev}, \citenamefont {Wu}, \citenamefont {Armbruster}, \citenamefont
  {Naghilou}, \citenamefont {Brouwer}, \citenamefont {Ivanov}, \citenamefont
  {Derrien}, \citenamefont {Bulgakova}, \citenamefont {Kautek}, \citenamefont
  {Rethfeld} \emph {et~al.}}]{shugaev2016fundamentals}%
  \BibitemOpen
  \bibfield  {author} {\bibinfo {author} {\bibfnamefont {M.~V.}\ \bibnamefont
  {Shugaev}}, \bibinfo {author} {\bibfnamefont {C.}~\bibnamefont {Wu}},
  \bibinfo {author} {\bibfnamefont {O.}~\bibnamefont {Armbruster}}, \bibinfo
  {author} {\bibfnamefont {A.}~\bibnamefont {Naghilou}}, \bibinfo {author}
  {\bibfnamefont {N.}~\bibnamefont {Brouwer}}, \bibinfo {author} {\bibfnamefont
  {D.~S.}\ \bibnamefont {Ivanov}}, \bibinfo {author} {\bibfnamefont {T.~J.-Y.}\
  \bibnamefont {Derrien}}, \bibinfo {author} {\bibfnamefont {N.~M.}\
  \bibnamefont {Bulgakova}}, \bibinfo {author} {\bibfnamefont {W.}~\bibnamefont
  {Kautek}}, \bibinfo {author} {\bibfnamefont {B.}~\bibnamefont {Rethfeld}},
  \emph {et~al.},\ }\bibfield  {title} {\bibinfo {title} {Fundamentals of
  ultrafast laser--material interaction},\ }\href@noop {} {\bibfield  {journal}
  {\bibinfo  {journal} {Mrs Bulletin}\ }\textbf {\bibinfo {volume} {41}},\
  \bibinfo {pages} {960} (\bibinfo {year} {2016})}\BibitemShut {NoStop}%
\bibitem [{\citenamefont {Rethfeld}\ \emph {et~al.}(2017)\citenamefont
  {Rethfeld}, \citenamefont {Ivanov}, \citenamefont {Garcia},\ and\
  \citenamefont {Anisimov}}]{Rethfeld2017}%
  \BibitemOpen
  \bibfield  {author} {\bibinfo {author} {\bibfnamefont {B.}~\bibnamefont
  {Rethfeld}}, \bibinfo {author} {\bibfnamefont {D.~S.}\ \bibnamefont
  {Ivanov}}, \bibinfo {author} {\bibfnamefont {M.~E.}\ \bibnamefont {Garcia}},\
  and\ \bibinfo {author} {\bibfnamefont {S.~I.}\ \bibnamefont {Anisimov}},\
  }\bibfield  {title} {\bibinfo {title} {Modelling ultrafast laser ablation},\
  }\href {https://doi.org/10.1088/1361-6463/50/19/193001} {\bibfield  {journal}
  {\bibinfo  {journal} {Journal of Physics D: Applied Physics}\ }\textbf
  {\bibinfo {volume} {50}},\ \bibinfo {pages} {193001} (\bibinfo {year}
  {2017})}\BibitemShut {NoStop}%
\bibitem [{\citenamefont {Anisimov}\ \emph {et~al.}(1974)\citenamefont
  {Anisimov}, \citenamefont {Kapeliovich},\ and\ \citenamefont
  {Perel'man}}]{Anisimov1974}%
  \BibitemOpen
  \bibfield  {author} {\bibinfo {author} {\bibfnamefont {S.~I.}\ \bibnamefont
  {Anisimov}}, \bibinfo {author} {\bibfnamefont {B.~L.}\ \bibnamefont
  {Kapeliovich}},\ and\ \bibinfo {author} {\bibfnamefont {T.~L.}\ \bibnamefont
  {Perel'man}},\ }\bibfield  {title} {\bibinfo {title} {{Electron emission from
  metal surfaces exposed to ultrashort laser pulses}},\ }\href
  {http://www.jetp.ac.ru/cgi-bin/e/index/e/39/2/p375?a=list} {\bibfield
  {journal} {\bibinfo  {journal} {Sov. Phys. JETP}\ }\textbf {\bibinfo {volume}
  {39}},\ \bibinfo {pages} {375} (\bibinfo {year} {1974})}\BibitemShut
  {NoStop}%
\bibitem [{\citenamefont {van Driel}(1987)}]{vanDriel1987}%
  \BibitemOpen
  \bibfield  {author} {\bibinfo {author} {\bibfnamefont {H.~M.}\ \bibnamefont
  {van Driel}},\ }\bibfield  {title} {\bibinfo {title} {{Kinetics of
  high-density plasmas generated in Si by {1.06- and 0.53-{\micro\metre}}
  picosecond laser pulses}},\ }\href {https://doi.org/10.1103/PhysRevB.35.8166}
  {\bibfield  {journal} {\bibinfo  {journal} {Phys. Rev. B}\ }\textbf {\bibinfo
  {volume} {35}},\ \bibinfo {pages} {8166} (\bibinfo {year}
  {1987})}\BibitemShut {NoStop}%
\bibitem [{\citenamefont {R\"{a}mer}\ \emph {et~al.}(2014)\citenamefont
  {R\"{a}mer}, \citenamefont {Osmani},\ and\ \citenamefont
  {Rethfeld}}]{Raemer2014}%
  \BibitemOpen
  \bibfield  {author} {\bibinfo {author} {\bibfnamefont {A.}~\bibnamefont
  {R\"{a}mer}}, \bibinfo {author} {\bibfnamefont {O.}~\bibnamefont {Osmani}},\
  and\ \bibinfo {author} {\bibfnamefont {B.}~\bibnamefont {Rethfeld}},\
  }\bibfield  {title} {\bibinfo {title} {{Laser damage in silicon: Energy
  absorption, relaxation, and transport}},\ }\href
  {https://doi.org/10.1063/1.4891633} {\bibfield  {journal} {\bibinfo
  {journal} {Journal of Applied Physics}\ }\textbf {\bibinfo {volume} {116}},\
  \bibinfo {pages} {053508} (\bibinfo {year} {2014})}\BibitemShut {NoStop}%
\bibitem [{\citenamefont {Ndione}\ \emph
  {et~al.}(2022{\natexlab{a}})\citenamefont {Ndione}, \citenamefont {Weber},
  \citenamefont {Gericke},\ and\ \citenamefont {Rethfeld}}]{Ndione2022}%
  \BibitemOpen
  \bibfield  {author} {\bibinfo {author} {\bibfnamefont {P.~D.}\ \bibnamefont
  {Ndione}}, \bibinfo {author} {\bibfnamefont {S.~T.}\ \bibnamefont {Weber}},
  \bibinfo {author} {\bibfnamefont {D.~O.}\ \bibnamefont {Gericke}},\ and\
  \bibinfo {author} {\bibfnamefont {B.}~\bibnamefont {Rethfeld}},\ }\bibfield
  {title} {\bibinfo {title} {Nonequilibrium band occupation and optical
  response of gold after ultrafast xuv excitation},\ }\href@noop {} {\bibfield
  {journal} {\bibinfo  {journal} {Scientific Reports}\ }\textbf {\bibinfo
  {volume} {12}},\ \bibinfo {pages} {1} (\bibinfo {year}
  {2022}{\natexlab{a}})}\BibitemShut {NoStop}%
\bibitem [{\citenamefont {Ndione}\ \emph
  {et~al.}(2022{\natexlab{b}})\citenamefont {Ndione}, \citenamefont {Gericke},\
  and\ \citenamefont {Rethfeld}}]{Ndione2022b}%
  \BibitemOpen
  \bibfield  {author} {\bibinfo {author} {\bibfnamefont {P.~D.}\ \bibnamefont
  {Ndione}}, \bibinfo {author} {\bibfnamefont {D.~O.}\ \bibnamefont
  {Gericke}},\ and\ \bibinfo {author} {\bibfnamefont {B.}~\bibnamefont
  {Rethfeld}},\ }\bibfield  {title} {\bibinfo {title} {Optical properties of
  gold after intense short-pulse excitations},\ }\bibfield  {journal} {\bibinfo
   {journal} {Frontiers in Physics}\ }\textbf {\bibinfo {volume} {10}},\ \href
  {https://doi.org/10.3389/fphy.2022.856817} {10.3389/fphy.2022.856817}
  (\bibinfo {year} {2022}{\natexlab{b}})\BibitemShut {NoStop}%
\bibitem [{\citenamefont {Beaurepaire}\ \emph {et~al.}(1996)\citenamefont
  {Beaurepaire}, \citenamefont {Merle}, \citenamefont {Daunois},\ and\
  \citenamefont {Bigot}}]{Beaurepaire1996}%
  \BibitemOpen
  \bibfield  {author} {\bibinfo {author} {\bibfnamefont {E.}~\bibnamefont
  {Beaurepaire}}, \bibinfo {author} {\bibfnamefont {J.-C.}\ \bibnamefont
  {Merle}}, \bibinfo {author} {\bibfnamefont {A.}~\bibnamefont {Daunois}},\
  and\ \bibinfo {author} {\bibfnamefont {J.-Y.}\ \bibnamefont {Bigot}},\
  }\bibfield  {title} {\bibinfo {title} {Ultrafast spin dynamics in
  ferromagnetic nickel},\ }\href {https://doi.org/10.1103/PhysRevLett.76.4250}
  {\bibfield  {journal} {\bibinfo  {journal} {Physical Review Letters}\
  }\textbf {\bibinfo {volume} {76}},\ \bibinfo {pages} {4250} (\bibinfo {year}
  {1996})}\BibitemShut {NoStop}%
\bibitem [{\citenamefont {Koopmans}\ \emph {et~al.}(2000)\citenamefont
  {Koopmans}, \citenamefont {van Kampen}, \citenamefont {Kohlhepp},\ and\
  \citenamefont {de~Jonge}}]{Koopmans2000}%
  \BibitemOpen
  \bibfield  {author} {\bibinfo {author} {\bibfnamefont {B.}~\bibnamefont
  {Koopmans}}, \bibinfo {author} {\bibfnamefont {M.}~\bibnamefont {van
  Kampen}}, \bibinfo {author} {\bibfnamefont {J.~T.}\ \bibnamefont
  {Kohlhepp}},\ and\ \bibinfo {author} {\bibfnamefont {W.~J.~M.}\ \bibnamefont
  {de~Jonge}},\ }\bibfield  {title} {\bibinfo {title} {Ultrafast magneto-optics
  in nickel: Magnetism or optics?},\ }\href
  {https://doi.org/10.1103/PhysRevLett.85.844} {\bibfield  {journal} {\bibinfo
  {journal} {Phys. Rev. Lett.}\ }\textbf {\bibinfo {volume} {85}},\ \bibinfo
  {pages} {844} (\bibinfo {year} {2000})}\BibitemShut {NoStop}%
\bibitem [{\citenamefont {Mueller}\ and\ \citenamefont
  {Rethfeld}(2014)}]{Mueller2014PRB}%
  \BibitemOpen
  \bibfield  {author} {\bibinfo {author} {\bibfnamefont {B.~Y.}\ \bibnamefont
  {Mueller}}\ and\ \bibinfo {author} {\bibfnamefont {B.}~\bibnamefont
  {Rethfeld}},\ }\bibfield  {title} {\bibinfo {title} {{Thermodynamic
  $\ensuremath{\mu}T$ model of ultrafast magnetization dynamics}},\ }\href
  {https://doi.org/10.1103/PhysRevB.90.144420} {\bibfield  {journal} {\bibinfo
  {journal} {Phys. Rev. B}\ }\textbf {\bibinfo {volume} {90}},\ \bibinfo
  {pages} {144420} (\bibinfo {year} {2014})}\BibitemShut {NoStop}%
\bibitem [{\citenamefont {Waldecker}\ \emph {et~al.}(2016)\citenamefont
  {Waldecker}, \citenamefont {Bertoni}, \citenamefont {Ernstorfer},\ and\
  \citenamefont {Vorberger}}]{Waldecker2016}%
  \BibitemOpen
  \bibfield  {author} {\bibinfo {author} {\bibfnamefont {L.}~\bibnamefont
  {Waldecker}}, \bibinfo {author} {\bibfnamefont {R.}~\bibnamefont {Bertoni}},
  \bibinfo {author} {\bibfnamefont {R.}~\bibnamefont {Ernstorfer}},\ and\
  \bibinfo {author} {\bibfnamefont {J.}~\bibnamefont {Vorberger}},\ }\bibfield
  {title} {\bibinfo {title} {Electron-phonon coupling and energy flow in a
  simple metal beyond the two-temperature approximation},\ }\href
  {https://doi.org/10.1103/PhysRevX.6.021003} {\bibfield  {journal} {\bibinfo
  {journal} {Phys. Rev. X}\ }\textbf {\bibinfo {volume} {6}},\ \bibinfo {pages}
  {021003} (\bibinfo {year} {2016})}\BibitemShut {NoStop}%
\bibitem [{\citenamefont {Carpene}(2006)}]{Carpene2006}%
  \BibitemOpen
  \bibfield  {author} {\bibinfo {author} {\bibfnamefont {E.}~\bibnamefont
  {Carpene}},\ }\bibfield  {title} {\bibinfo {title} {Ultrafast laser
  irradiation of metals: Beyond the two-temperature model},\ }\href
  {https://doi.org/10.1103/PhysRevB.74.024301} {\bibfield  {journal} {\bibinfo
  {journal} {Phys. Rev. B}\ }\textbf {\bibinfo {volume} {74}},\ \bibinfo
  {pages} {024301} (\bibinfo {year} {2006})}\BibitemShut {NoStop}%
\bibitem [{\citenamefont {Tsibidis}(2018)}]{Tsibidis2018}%
  \BibitemOpen
  \bibfield  {author} {\bibinfo {author} {\bibfnamefont {G.~D.}\ \bibnamefont
  {Tsibidis}},\ }\bibfield  {title} {\bibinfo {title} {Ultrafast dynamics of
  non-equilibrium electrons and strain generation under femtosecond laser
  irradiation of nickel},\ }\bibfield  {journal} {\bibinfo  {journal} {Applied
  Physics A}\ }\textbf {\bibinfo {volume} {124}},\ \href
  {https://doi.org/10.1007/s00339-018-1704-4} {10.1007/s00339-018-1704-4}
  (\bibinfo {year} {2018})\BibitemShut {NoStop}%
\bibitem [{\citenamefont {Uehlein}\ \emph {et~al.}(2022)\citenamefont
  {Uehlein}, \citenamefont {Weber},\ and\ \citenamefont
  {Rethfeld}}]{Uehlein2022}%
  \BibitemOpen
  \bibfield  {author} {\bibinfo {author} {\bibfnamefont {M.}~\bibnamefont
  {Uehlein}}, \bibinfo {author} {\bibfnamefont {S.~T.}\ \bibnamefont {Weber}},\
  and\ \bibinfo {author} {\bibfnamefont {B.}~\bibnamefont {Rethfeld}},\
  }\bibfield  {title} {\bibinfo {title} {Influence of electronic
  non-equilibrium on energy distribution and dissipation in aluminum studied
  with an extended two-temperature model},\ }\href
  {https://doi.org/10.3390/nano12101655} {\bibfield  {journal} {\bibinfo
  {journal} {Nanomaterials}\ }\textbf {\bibinfo {volume} {12}},\ \bibinfo
  {pages} {1655} (\bibinfo {year} {2022})}\BibitemShut {NoStop}%
\bibitem [{\citenamefont {Miyamoto}(2021)}]{Miyamoto2021}%
  \BibitemOpen
  \bibfield  {author} {\bibinfo {author} {\bibfnamefont {Y.}~\bibnamefont
  {Miyamoto}},\ }\bibfield  {title} {\bibinfo {title} {{Direct treatment of
  interaction between laser-field and electrons for simulating laser processing
  of metals}},\ }\href {https://doi.org/10.1038/s41598-021-94036-4} {\bibfield
  {journal} {\bibinfo  {journal} {Scientific Reports}\ }\textbf {\bibinfo
  {volume} {11}},\ \bibinfo {pages} {14626} (\bibinfo {year}
  {2021})}\BibitemShut {NoStop}%
\bibitem [{\citenamefont {Yamada}\ and\ \citenamefont
  {Yabana}(2019)}]{Yamada2019}%
  \BibitemOpen
  \bibfield  {author} {\bibinfo {author} {\bibfnamefont {A.}~\bibnamefont
  {Yamada}}\ and\ \bibinfo {author} {\bibfnamefont {K.}~\bibnamefont
  {Yabana}},\ }\bibfield  {title} {\bibinfo {title} {{Energy transfer from
  intense laser pulse to dielectrics in time-dependent density functional
  theory}},\ }\href {https://doi.org/10.1140/epjd/e2019-90334-7} {\bibfield
  {journal} {\bibinfo  {journal} {The European Physical Journal D}\ }\textbf
  {\bibinfo {volume} {73}},\ \bibinfo {pages} {87} (\bibinfo {year}
  {2019})}\BibitemShut {NoStop}%
\bibitem [{\citenamefont {Kaiser}\ \emph {et~al.}(2000)\citenamefont {Kaiser},
  \citenamefont {Rethfeld}, \citenamefont {Vicanek},\ and\ \citenamefont
  {Simon}}]{Kaiser2000}%
  \BibitemOpen
  \bibfield  {author} {\bibinfo {author} {\bibfnamefont {A.}~\bibnamefont
  {Kaiser}}, \bibinfo {author} {\bibfnamefont {B.}~\bibnamefont {Rethfeld}},
  \bibinfo {author} {\bibfnamefont {M.}~\bibnamefont {Vicanek}},\ and\ \bibinfo
  {author} {\bibfnamefont {G.}~\bibnamefont {Simon}},\ }\bibfield  {title}
  {\bibinfo {title} {Microscopic processes in dielectrics under irradiation by
  subpicosecond laser pulses},\ }\href
  {https://doi.org/10.1103/PhysRevB.61.11437} {\bibfield  {journal} {\bibinfo
  {journal} {Phys. Rev. B}\ }\textbf {\bibinfo {volume} {61}},\ \bibinfo
  {pages} {11437} (\bibinfo {year} {2000})}\BibitemShut {NoStop}%
\bibitem [{\citenamefont {Brouwer}\ and\ \citenamefont
  {Rethfeld}(2014)}]{Brouwer2014}%
  \BibitemOpen
  \bibfield  {author} {\bibinfo {author} {\bibfnamefont {N.}~\bibnamefont
  {Brouwer}}\ and\ \bibinfo {author} {\bibfnamefont {B.}~\bibnamefont
  {Rethfeld}},\ }\bibfield  {title} {\bibinfo {title} {Excitation and
  relaxation dynamics in dielectrics irradiated by an intense ultrashort laser
  pulse},\ }\href {https://doi.org/10.1364/JOSAB.31.000C28} {\bibfield
  {journal} {\bibinfo  {journal} {J. Opt. Soc. Am. B}\ }\textbf {\bibinfo
  {volume} {31}},\ \bibinfo {pages} {C28} (\bibinfo {year} {2014})}\BibitemShut
  {NoStop}%
\bibitem [{\citenamefont {Brouwer}\ and\ \citenamefont
  {Rethfeld}(2017)}]{brouwer2017}%
  \BibitemOpen
  \bibfield  {author} {\bibinfo {author} {\bibfnamefont {N.}~\bibnamefont
  {Brouwer}}\ and\ \bibinfo {author} {\bibfnamefont {B.}~\bibnamefont
  {Rethfeld}},\ }\bibfield  {title} {\bibinfo {title} {Transient electron
  excitation and nonthermal electron-phonon coupling in dielectrics irradiated
  by ultrashort laser pulses},\ }\href
  {https://journals.aps.org/prb/abstract/10.1103/PhysRevB.95.245139} {\bibfield
   {journal} {\bibinfo  {journal} {Physical Review B}\ }\textbf {\bibinfo
  {volume} {95}},\ \bibinfo {pages} {245139} (\bibinfo {year}
  {2017})}\BibitemShut {NoStop}%
\bibitem [{\citenamefont {Brouwer}(2017)}]{phd_Brouwer}%
  \BibitemOpen
  \bibfield  {author} {\bibinfo {author} {\bibfnamefont {N.}~\bibnamefont
  {Brouwer}},\ }\emph {\bibinfo {title} {Non-equilibrium electron dynamics in
  dielectrics irradiated by ultrashort laser pulses}},\ \href@noop {} {\bibinfo
  {type} {Doktorarbeit}},\ \bibinfo  {school} {Technische Universit\"{a}t
  Kaiserslautern} (\bibinfo {year} {2017})\BibitemShut {NoStop}%
\bibitem [{\citenamefont {Rethfeld}(2004)}]{Rethfeld2004a}%
  \BibitemOpen
  \bibfield  {author} {\bibinfo {author} {\bibfnamefont {B.}~\bibnamefont
  {Rethfeld}},\ }\bibfield  {title} {\bibinfo {title} {Unified model for the
  free-electron avalanche in laser-irradiated dielectrics},\ }\href
  {https://doi.org/10.1103/PhysRevLett.92.187401} {\bibfield  {journal}
  {\bibinfo  {journal} {Phys. Rev. Lett.}\ }\textbf {\bibinfo {volume} {92}},\
  \bibinfo {pages} {187401} (\bibinfo {year} {2004})}\BibitemShut {NoStop}%
\bibitem [{\citenamefont {Klett}\ and\ \citenamefont
  {Rethfeld}(2018)}]{Klett2018}%
  \BibitemOpen
  \bibfield  {author} {\bibinfo {author} {\bibfnamefont {I.}~\bibnamefont
  {Klett}}\ and\ \bibinfo {author} {\bibfnamefont {B.}~\bibnamefont
  {Rethfeld}},\ }\bibfield  {title} {\bibinfo {title} {Relaxation of a
  nonequilibrium phonon distribution induced by femtosecond laser
  irradiation},\ }\href {https://doi.org/10.1103/PhysRevB.98.144306} {\bibfield
   {journal} {\bibinfo  {journal} {Phys. Rev. B}\ }\textbf {\bibinfo {volume}
  {98}},\ \bibinfo {pages} {144306} (\bibinfo {year} {2018})}\BibitemShut
  {NoStop}%
\bibitem [{\citenamefont {Caruso}\ and\ \citenamefont
  {Novko}(2022)}]{Caruso2022}%
  \BibitemOpen
  \bibfield  {author} {\bibinfo {author} {\bibfnamefont {F.}~\bibnamefont
  {Caruso}}\ and\ \bibinfo {author} {\bibfnamefont {D.}~\bibnamefont {Novko}},\
  }\bibfield  {title} {\bibinfo {title} {Ultrafast dynamics of electrons and
  phonons: from the two-temperature model to the time-dependent boltzmann
  equation},\ }\href {https://doi.org/10.1080/23746149.2022.2095925} {\bibfield
   {journal} {\bibinfo  {journal} {Advances in Physics: X}\ }\textbf {\bibinfo
  {volume} {7}},\ \bibinfo {pages} {2095925} (\bibinfo {year} {2022})},\
  \Eprint {https://arxiv.org/abs/https://doi.org/10.1080/23746149.2022.2095925}
  {https://doi.org/10.1080/23746149.2022.2095925} \BibitemShut {NoStop}%
\bibitem [{\citenamefont {Ridley}(1999)}]{Ridley1999}%
  \BibitemOpen
  \bibfield  {author} {\bibinfo {author} {\bibfnamefont {B.~K.}\ \bibnamefont
  {Ridley}},\ }\href@noop {} {\emph {\bibinfo {title} {Quantum Processes in
  Semiconductors}}}\ (\bibinfo  {publisher} {Oxford University Press},\
  \bibinfo {year} {1999})\BibitemShut {NoStop}%
\bibitem [{\citenamefont {Galassi}\ \emph {et~al.}(2009)\citenamefont
  {Galassi}, \citenamefont {Davies}, \citenamefont {Theiler}, \citenamefont
  {Gough}, \citenamefont {Jungman}, \citenamefont {Alken}, \citenamefont
  {Booth},\ and\ \citenamefont {Rossi}}]{gsl}%
  \BibitemOpen
  \bibfield  {author} {\bibinfo {author} {\bibfnamefont {M.}~\bibnamefont
  {Galassi}}, \bibinfo {author} {\bibfnamefont {J.}~\bibnamefont {Davies}},
  \bibinfo {author} {\bibfnamefont {J.}~\bibnamefont {Theiler}}, \bibinfo
  {author} {\bibfnamefont {B.}~\bibnamefont {Gough}}, \bibinfo {author}
  {\bibfnamefont {G.}~\bibnamefont {Jungman}}, \bibinfo {author} {\bibfnamefont
  {P.}~\bibnamefont {Alken}}, \bibinfo {author} {\bibfnamefont
  {M.}~\bibnamefont {Booth}},\ and\ \bibinfo {author} {\bibfnamefont
  {F.}~\bibnamefont {Rossi}},\ }\href {http://www.gnu.org/software/gsl} {\emph
  {\bibinfo {title} {{GNU Scientific Library Reference Manual}}}},\ \bibinfo
  {edition} {3rd}\ ed.\ (\bibinfo  {publisher} {Network Theory Ltd},\ \bibinfo
  {year} {2009})\BibitemShut {NoStop}%
\bibitem [{\citenamefont {Rethfeld}\ \emph {et~al.}(2002)\citenamefont
  {Rethfeld}, \citenamefont {Kaiser}, \citenamefont {Vicanek},\ and\
  \citenamefont {Simon}}]{Rethfeld2002}%
  \BibitemOpen
  \bibfield  {author} {\bibinfo {author} {\bibfnamefont {B.}~\bibnamefont
  {Rethfeld}}, \bibinfo {author} {\bibfnamefont {A.}~\bibnamefont {Kaiser}},
  \bibinfo {author} {\bibfnamefont {M.}~\bibnamefont {Vicanek}},\ and\ \bibinfo
  {author} {\bibfnamefont {G.}~\bibnamefont {Simon}},\ }\bibfield  {title}
  {\bibinfo {title} {Ultrafast dynamics of nonequilibrium electrons in metals
  under femtosecond laser irradiation},\ }\href
  {https://doi.org/10.1103/PhysRevB.65.214303} {\bibfield  {journal} {\bibinfo
  {journal} {Phys. Rev. B}\ }\textbf {\bibinfo {volume} {65}},\ \bibinfo
  {pages} {214303} (\bibinfo {year} {2002})}\BibitemShut {NoStop}%
\bibitem [{\citenamefont {Medvedev}\ \emph {et~al.}(2011)\citenamefont
  {Medvedev}, \citenamefont {Zastrau}, \citenamefont {F\"orster}, \citenamefont
  {Gericke},\ and\ \citenamefont {Rethfeld}}]{Medvedev2011}%
  \BibitemOpen
  \bibfield  {author} {\bibinfo {author} {\bibfnamefont {N.}~\bibnamefont
  {Medvedev}}, \bibinfo {author} {\bibfnamefont {U.}~\bibnamefont {Zastrau}},
  \bibinfo {author} {\bibfnamefont {E.}~\bibnamefont {F\"orster}}, \bibinfo
  {author} {\bibfnamefont {D.~O.}\ \bibnamefont {Gericke}},\ and\ \bibinfo
  {author} {\bibfnamefont {B.}~\bibnamefont {Rethfeld}},\ }\bibfield  {title}
  {\bibinfo {title} {Short-time electron dynamics in aluminum excited by
  femtosecond extreme ultraviolet radiation},\ }\href
  {https://doi.org/10.1103/PhysRevLett.107.165003} {\bibfield  {journal}
  {\bibinfo  {journal} {Phys. Rev. Lett.}\ }\textbf {\bibinfo {volume} {107}},\
  \bibinfo {pages} {165003} (\bibinfo {year} {2011})}\BibitemShut {NoStop}%
\bibitem [{\citenamefont {Landau}\ and\ \citenamefont
  {Lifschitz}(1980)}]{landau1980lifshitz}%
  \BibitemOpen
  \bibfield  {author} {\bibinfo {author} {\bibfnamefont {L.~D.}\ \bibnamefont
  {Landau}}\ and\ \bibinfo {author} {\bibfnamefont {E.~M.}\ \bibnamefont
  {Lifschitz}},\ }\href@noop {} {\emph {\bibinfo {title} {Course Of Theoretical
  Physics}}},\ Vol.\ \bibinfo {volume} {5: Statistical Physics Part 1}\
  (\bibinfo  {publisher} {Pergamon Press, Oxford, New York},\ \bibinfo {year}
  {1980})\BibitemShut {NoStop}%
\end{thebibliography}%

\end{document}